\documentclass[aps,nofootinbib,preprintnumbers,showpacs,twocolumn,prd]{revtex4-1}

\usepackage{mathrsfs}
\usepackage{graphicx}
\usepackage{float}
\usepackage{bm}
\usepackage{hyperref}
\usepackage{amsmath}
\usepackage{amssymb}
\begin{document}

\title{Boer-Mulders function of pion meson and $q_T$-weighted $\cos2\phi$ asymmetry in the unpolarized $\pi^- p$ Drell-Yan at COMPASS}
\author{Zhengxian Wang$^\ast$}
\affiliation{Department of Physics, Southeast University, Nanjing 211189, China}
\author{Xiaoyu Wang}
\thanks{These authors contributed equally to this work and should be considered as co-first authors}
\affiliation{Department of Physics, Southeast University, Nanjing 211189, China}
\author{Zhun Lu}
\email{E-mail:zhunlu@seu.edu.cn}
\affiliation{Department of Physics, Southeast University, Nanjing 211189, China}

\begin{abstract}
We calculate the two leading-twist transverse momentum dependent distribution functions of the pion meson, the unpolarized distribution $f_{1\pi}(x,\bm{k}^2_T)$ and the Boer-Mulders function $h_{1\pi}^\perp (x,\bm{k}^2_T)$, using the pion wave functions derived from a light-cone approach. The evolution effect of the first $\bm k_T$-moment of the pion Boer-Mulders function is studied by employing an approximate evolution kernel. Using the model resulting distributions, we predict the transverse momentum weighted $\cos2\phi$ azimuthal asymmetry in the unpolarized $\pi^- p$ Drell-Yan process which can be measured at COMPASS in the near future.
\end{abstract}

\maketitle

\section{Introduction}
\label{Sec.intro}

The Boer-Mulders function is a transverse momentum dependent (TMD) distribution function which describes the transversely polarization distribution of the quark inside an unpolarized hadron~\cite{Boer:1997nt}.
It has attracted a lot of attention because it manifests novel structure of hadron that encodes the correlation of quark transverse momentum and transverse spin~\cite{Lu:2016pdp}.
It also brings new insights to QCD dynamics as a nonvanishing Boer-Mulders function requires gluon rescattering effect~\cite{Brodsky:2002cx}.
In this case the gauge-link in the operator definition of TMD distributions plays an essential role~\cite{Collins:2002kn,Ji:2002aa,Belitsky:2002sm,Boer:2003cm}.
Furthermore, the convolution of two Boer-Mulders functions can provide leading-twist contribution~\cite{Boer:1999mm} to the $\cos 2\phi$ angular asymmetry observed in unpolarized Drell-Yan processes, which may account for the violation of the Lam-Tung relation~\cite{Lam:1978pu}.
For these reasons, the Boer-Mulders function of the proton has been studied extensively in literature by several QCD-inspired quark models, such as the spectator model~\cite{Boer:2002ju,Gamberg:2003ey,Bacchetta:2003rz,Lu:2006ew,Gamberg:2007wm,Burkardt:2007xm,Bacchetta:2008af}, the large $N_c$ model~\cite{Pobylitsa:2003ty}, the bag model~\cite{Yuan:2003wk,Courtoy:2009pc} and the (light-front) constituent quark model~\cite{Courtoy:2009pc,Pasquini:2010af}.
A considerable volume of phenomenological work focusing on the Boer-Mulders effect has been proposed to understand the $\cos2\phi$ asymmetry in unpolarized Drell-Yan processes~\cite{Boer:2002ju,lu_04,bianconi_04,Lu:2005rq,bianconi_05a,sissakian_05a,gamberg_05,
sissakian_05b,Lu:2006ew,barone_06,lu_07a,reimer_07,miller_07,bianconi_08,
sissakian_08,Lu:2011mz,Liu:2012fha,Liu:2012vn, Pasquini:2014ppa}.
Besides, several parameterizations~\cite{Zhang:2008nu,Lu:2009ip,Barone:2009hw,Barone:2010gk} of the proton Boer-Mulders function have been performed based on the measurements of the unpolarized $pp$ and $pd$ Drell-Yan processes~\cite{Zhu:2006gx,Zhu:2008sj} and the semi-inclusive deeply inelastic process(SIDIS)~\cite{Kafer:2008ud,Bressan:2009eu,Giordano:2009hi,Airapetian:2012yg}.

In recent years, the model calculation of the Boer-Mulders function has also been extended to the case of the pion meson.
The first model applied to calculate the Boer-Mulders function of the pion is the spectator model~\cite{Lu:2004au,Meissner:2008ay,Gamberg:2009uk}, in which the antiquark plays the role of the spectator particle.
In Refs.~\cite{Lu:2004au} and \cite{Meissner:2008ay}, the gauge-link was taken into account by the one-gluon exchange approximation, while in Ref.~\cite{Gamberg:2009uk}, the authors included higher-order gluonic contributions from the gauge link by applying non-perturbative eikonal methods.
Recently, the Boer-Mulders function of the pion was also calculated by the light-front constituent quark model~\cite{Pasquini:2014ppa} and the bag model~\cite{Lu:2012hh}.

As the pion meson is an unstable particle, it can not be served as a target in the SIDIS process.
A suitable approach to study the partonic structure of pion is the $\pi N$ Drell-Yan process in which a charged $\pi$ beam collides on a nucleon target.
This idea was exploited decades ago by the NA10 Collaboration~\cite{na10} and the E615 Collaboration~\cite{conway}, which measured the angular asymmetries in the process $\pi^-\, N\rightarrow \mu^+\mu^- \,X$, with $N$ denoting a nucleon in the deuterium or tungsten target.
Recently, new pion-induced Drell-Yan program which can be conducted at the COMPASS facility was proposed~\cite{Gautheron:2010wva}, and the first Drell-Yan data were taken using a high-intensity $\pi^-$ beam of 190 GeV and a transversely polarised proton target~\cite{Adolph:2016dvl}.
The Drell-Yan program at COMPASS also provides great opportunity to explore the Boer-Mulders function of the pion via employing an unpolarized target or averaging the polarized data.
Therefore, in this work, we will study the $\cos2\phi$ asymmetry contributed by the Boer-Mulders function at the kinematics of COMPASS.
To do this we calculate the Boer-Mulders function of the pion using a light-cone approach, in which distribution functions can be expressed as an overlap integration of the light-cone wave functions of the hadron.
In the calculation we adopt the pion wave function obtained from a light-cone quark model~\cite{Xiao:2003wf} after considering the relativistic effect~\cite{Melosh:1974cu,Ma:1991xq}.
Moreover, we take into account the $Q^2$ evolution effect of the first $\bm k_T$-moment of the Boer-Mulders function in order to present a more reliable prediction on $\cos2\phi$ asymmetry weighted by the transverse momentum of the dilepton $q_T$.

In this work we will only consider the contribution of the Boer-Mulders effect to the $\cos2\phi$ asymmetry in fixed-target $\pi p$ Drell-Yan.
However, a recent study~\cite{Lambertsen:2016wgj} based on NNLO perturbative-QCD (pQCD) shows that pQCD collinear effects seem to account for the violation of the Lam-Tung relation, especially at colliders~\cite{Aaltonen:2011nr,Khachatryan:2015paa}, but also in the fixed-target regime~\cite{Zhu:2006gx,Zhu:2008sj,na10,conway}.
This perhaps leaves little room for effects from intrinsic parton motion, which makes it more crucial to do more studies~\cite{Peng:2015spa} on the Drell-Yan $\cos 2\phi$ asymmetry.

The rest of the paper is organized as follows. In Sec.~II, we calculate the unpolarized distribution function of the pion meson using the pion wave functions derived from the light-cone approach. We determine the energy scale at which the model is valid by comparing the model calculation with the known parametrization of the unpolarized distribution function of the pion.
In Sec.~III, we employ the same model to calculate the pion Boer-Mulders function, using the overlap integration of the light-cone wave functions of the pion.
The evolution effect of $h_{1\pi}^{\perp (1)}(x)$ is also studied.
In Sec.~IV, we present our prediction on the transverse momentum weighted $\cos2\phi$ asymmetry in the unpolarized $\pi^- p$ Drell-Yan process at the kinematics of COMPASS.
We summarize our work in Sec.~V.

\section{Calculation on the unpolarized distribution function of the pion meson}

\label{Sec:unpolarized}

In this section, we present the model calculation of the unpolarized distribution function $f_1$ of the pion meson and compare the model calculated result with the known parametrization.
A convenient way to calculate the distribution functions is making use of the light-cone formalism~\cite{Brodsky:1997de}.
In this approach, the wave functions of the hadron, which describe a hadronic composite state at a particular light-cone time, are expressed in terms of a series of light-cone wave functions in Fock-state basis.
For example, the Fock states of the pion can be cast into
\begin{align}
|\pi\rangle = |q\bar{q}\rangle + |q\bar{q} g\rangle+\cdots~.
\end{align}
We only consider the first order contributions in the calculation to simplify the problem, ie., we take into account the minimal Fock states of the pion meson, which has been derived in Ref.~\cite{Xiao:2003wf} by considering the relativistic effect~\cite{Melosh:1974cu,Ma:1991xq}:
\begin{align}
\label{eq:LCWFs}
&\Psi_{\pi R}(x,\bm{k}_T,+,-)=+\frac{m}{\sqrt{2(m^2+\bm{k}^2_T)}}\varphi_\pi \qquad (l^z=0),\nonumber\\
&\Psi_{\pi R}(x,\bm{k}_T,-,+)=-\frac{m}{\sqrt{2(m^2+\bm{k}^2_T)}}\varphi_\pi \qquad (l^z=0),\nonumber\\
&\Psi_{\pi R}(x,\bm{k}_T,+,+)=-\frac{k_{T 1}-ik_{T 2}}{\sqrt{2(m^2+\bm{k}^2_T)}}\varphi_\pi \qquad (l^z=-1),\nonumber\\
&\Psi_{\pi R}(x,\bm{k}_T,-,-)=-\frac{k_{T 1}+ik_{T 2}}{\sqrt{2(m^2+\bm{k}^2_T)}}\varphi_\pi \qquad (l^z=+1).
\end{align}
Here, $x$ is the longitudinal momentum fraction of the quark in pion, $\bm{k}_T$ is the transverse momentum of the quark, $m$ stands for the mass of the quark/antiquark, and $+,-$ denotes the helicities of quark and the spectator antiquark, respectively.
$\varphi_\pi$ in Eq.~(\ref{eq:LCWFs}) is the wave function in momentum space, for which we adopt the Brodsky-Huang-Lepage (BHL) prescription~\cite{Brodsky:1980vj}:
\begin{align}
\label{eq:BHL}
\varphi_\pi(x,\bm{k}_T)=A\mathrm{exp}\left[-\frac{1}{8\beta^2}\frac{\bm{k}^2_T+m^2}{x(1-x)}\right].
\end{align}

\begin{figure*}
\centering
\scalebox{0.41}{\includegraphics*[1pt,0pt][596pt,419pt]{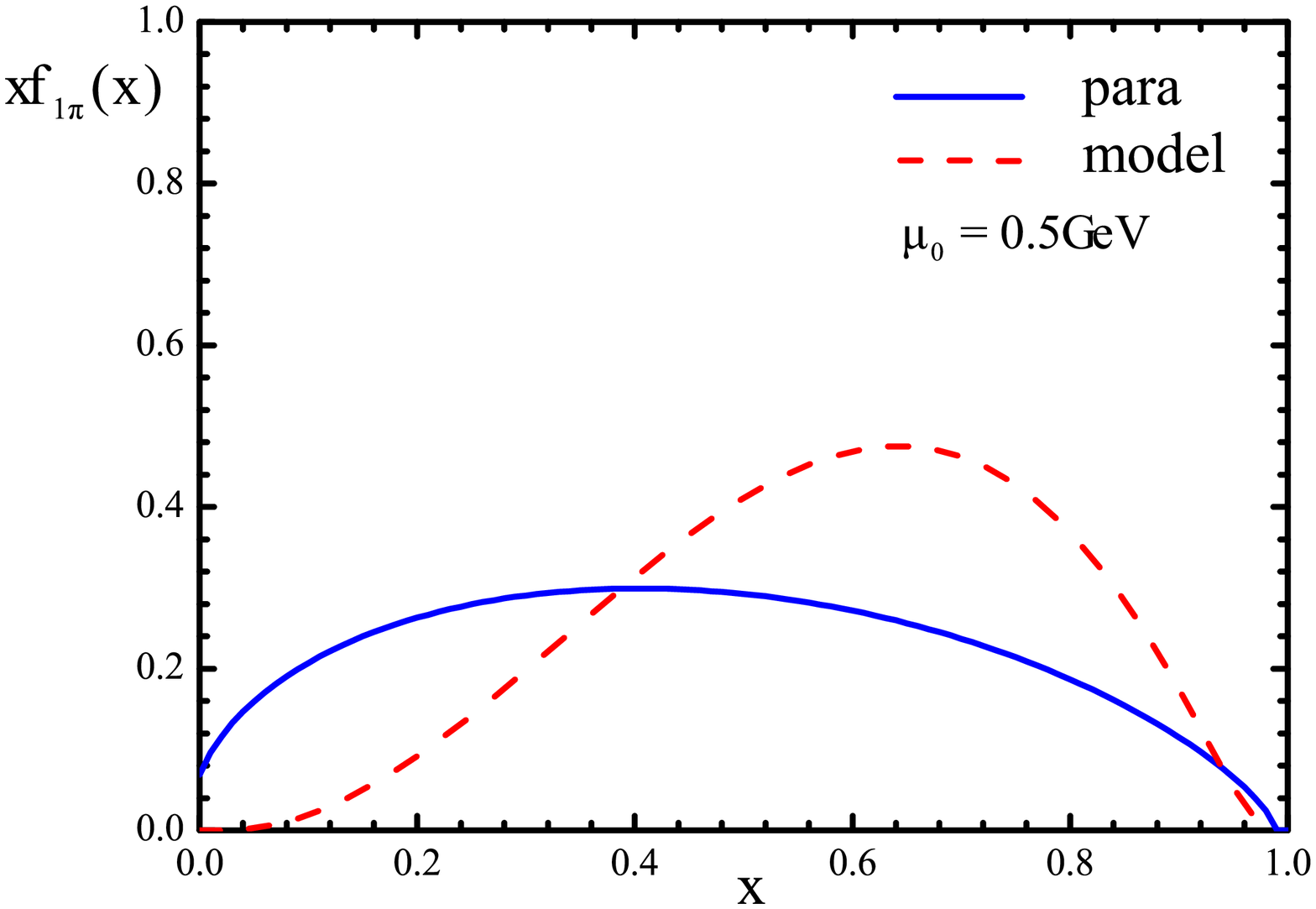}}
\scalebox{0.41}{\includegraphics*[1pt,0pt][599pt,421pt]{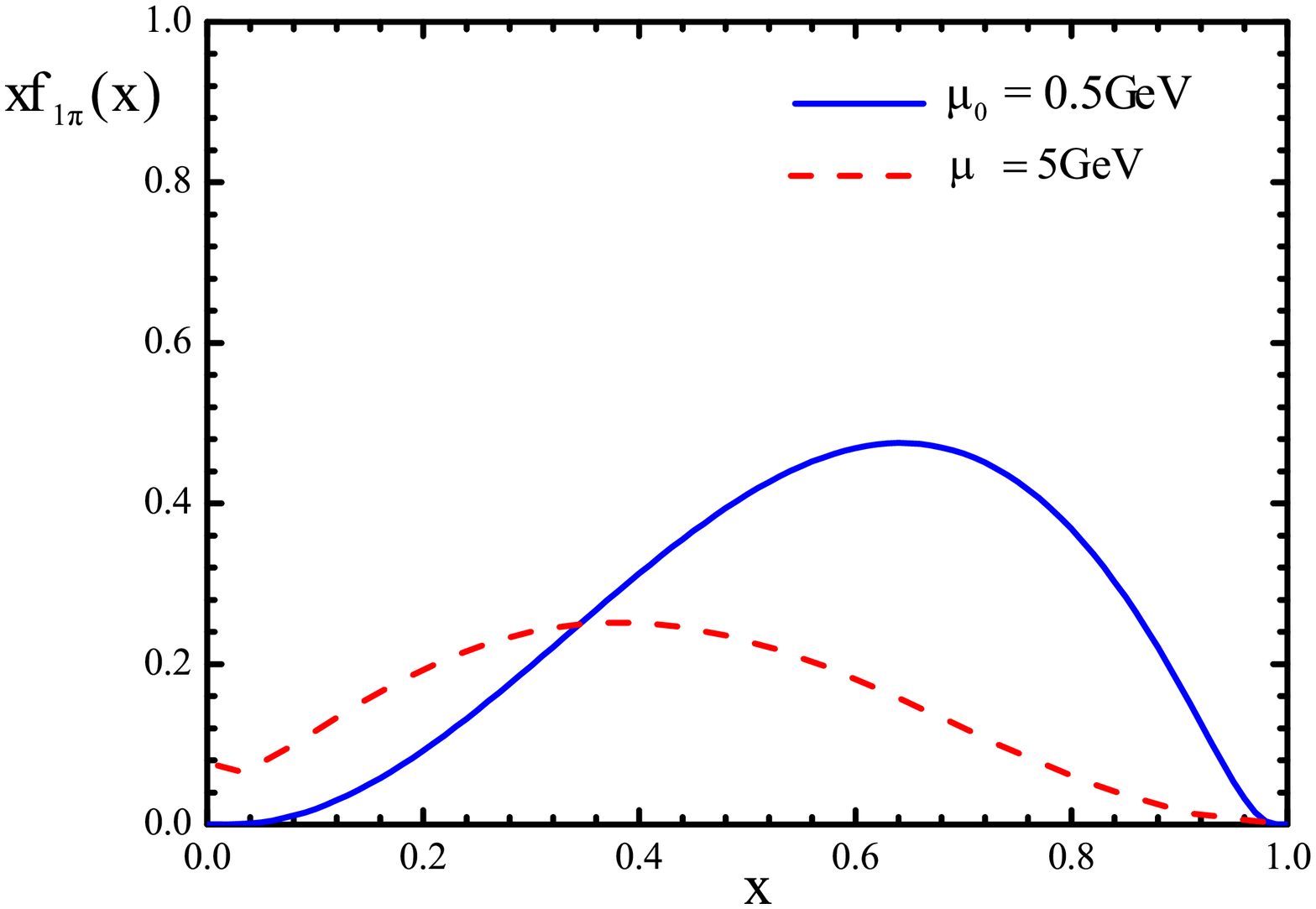}}
  \caption{Left panel: the unpolarized distribution function $xf_{1\pi}(x)$ vs $x$ from the model compared with the GRV LO parametrization of $xf_{1\pi}(x)$ at the same scale $\mu_0=0.5\ \mathrm{GeV}$. Right panel: comparison of $xf_{1\pi}(x)$ at two different scales $\mu_0=0.5\ \mathrm{GeV}$ and $\mu=5\ \mathrm{GeV}$.}
  \label{fig:f1}
\end{figure*}

The unpolarized quark TMD distribution can be written as
\begin{equation}
\label{eq:expansion}
{f}^{q/\pi}_{1}(x,\bm{k}_T^2)=\frac{1}{16{\pi}^{3}}\sum_{{\lambda }_{q}=\pm}\sum_{{\lambda }_{\bar{q}}=\pm}\left|{\Psi }_{{\lambda }_{q}{\lambda }_{\bar{q}}}\right|^2\,.
\end{equation}
Substituting Eqs.~(\ref{eq:LCWFs}),~(\ref{eq:BHL}) into Eq.~(\ref{eq:expansion}), we can obtain the expression of ${f}^{q}_{1}(x,\bm{k}_T^2)$
\begin{align}
\label{eq:unPDF}
&f_{1\pi}(x,\bm{k}_T^2)\nonumber\\&=\frac{1}{16\pi^3}\left(|\Psi_{++}|^2+|\Psi_{-+}|^2+
|\Psi_{+-}|^2+|\Psi_{--}|^2\right)\nonumber\\
&=\frac{1}{16\pi^3}A^2\mathrm{exp}\left[-\frac{1}{4\beta^2}\frac{\bm{k}_T^2+m^2}{x(1-x)}\right].
\end{align}
Here, we use $f_{1\pi}$ to denote the density of the valence quark in the charged pions:
\begin{align}
f_1^{\bar{u}/\pi^-} = f_1^{d/\pi^-} = f_1^{\bar{d}/\pi^+} = f_1^{u/\pi^+} \equiv  f_{1\pi}.
\end{align}
After integrating $f_{1\pi}(x,\bm{k}_T^2)$ over the transverse momentum $\bm{k}_T$, we can obtain the collinear distribution function
\begin{align}
\label{eq:unPDFx}
f_{1\pi}(x)&=\pi\int_0^\infty f_{1\pi}(x,\bm{k}^2_T)d\bm{k}^2_T\nonumber\\
&=\frac{A^2}{4\pi^2}\beta^2x(1-x)\mathrm{exp}\left[-\frac{1}{4\beta^2}\frac{m^2}{x(1-x)}\right].
\end{align}

To present the numerical result of $f_{1\pi}$, we need to specify the values of the parameters $A$, $m$ and $\beta$.
For the latter two we choose the values from Ref.~\cite{Xiao:2003wf}:
\begin{align}
&\beta=0.41\ \textrm{GeV},~~~
m_u=m_d=m=0.2\ \textrm{GeV}.\nonumber
\end{align}
In order to obtain an appropriate value for the parameter $A$, we compare the model result with the existing parameterization for $f_{1\pi}(x)$.
In our comparison we adopt the GRV LO parametrization~\cite{Gluck:1991ey} for $f_{1\pi}(x)$ at the lowest possible energy scale $\mu_0 = 0.5~\textrm{GeV}$, as we find that the lower possible scale is always preferred by the fit.
The fitted result is
\begin{align}
& A=31.303\ \textrm{GeV}^{-1}.
 \label{eq:modelscale}
\end{align}
Thus we assign $\mu_0 = 0.5$ GeV as the scale at which our model is applicable, the so-called model scale.
Once the model scale is specified, the results can be evolved to those at other energy scales in order to compare calculations with experimental measurements.

In the left panel of Fig~\ref{fig:f1}, we compare the valence quark distribution $f^{q_v}_{1\pi}(x,\mu)$ in the model (solid line) and in the GRV parametrization (dashed line) at the scale $\mu_0=0.5\ \mathrm{GeV}$.
On the other hand, we calculate the total momentum
carried by the valence quarks in the model and that in the GRV parametrization (in the case of $\pi^-$ meson), we find,
\begin{align}
\label{eq:compare}
&\int^{1}_{0}dx x[f^{\bar{u}_{v}}_{1\pi}(x,\mu_0)+f^{d_{v}}_{1\pi}(x,\mu_0)]_{\rm{model}} =0.50,\\
&\int^{1}_{0}dx x[f^{\bar{u}_{v}}_{1\pi}(x,\mu_0)+f^{d_{v}}_{1\pi}(x,\mu_0)]_{\rm{para}}=0.58.
\end{align}
The comparison shows that our model result is approximately consistent with the GRV parametrization.

In the right panel of Fig.~\ref{fig:f1}, we plot the valence quark distribution $f^{q_v}_{1\pi}(x,\mu)$ in our model at the model scale $\mu_0=0.5\  \mathrm{GeV}$ (solid line), as well as the evolved result of $f^{q_v}_{1\pi}(x,\mu)$ at the scale $\mu=5\ \mathrm{GeV}$ (dashed line).
To perform the DGLAP evolution on the model resulting PDF $f^{q}_{1\pi}(x,\mu)$, we adopt the {\sc{QCDNUM}}~\cite{Botje:2010ay} package at leading order, and we choose the strong coupling constant at the model scale as $\alpha_s(\mu_0^2)$= 0.911, which is obtained directly from the setting of the GRV parametrization.

\section{Model calculation of the Boer-mulders function}

\label{Sec:BMfunction}

\begin{figure*}
  \centering
\scalebox{0.41}{\includegraphics*[0pt,0pt][536pt,425pt]{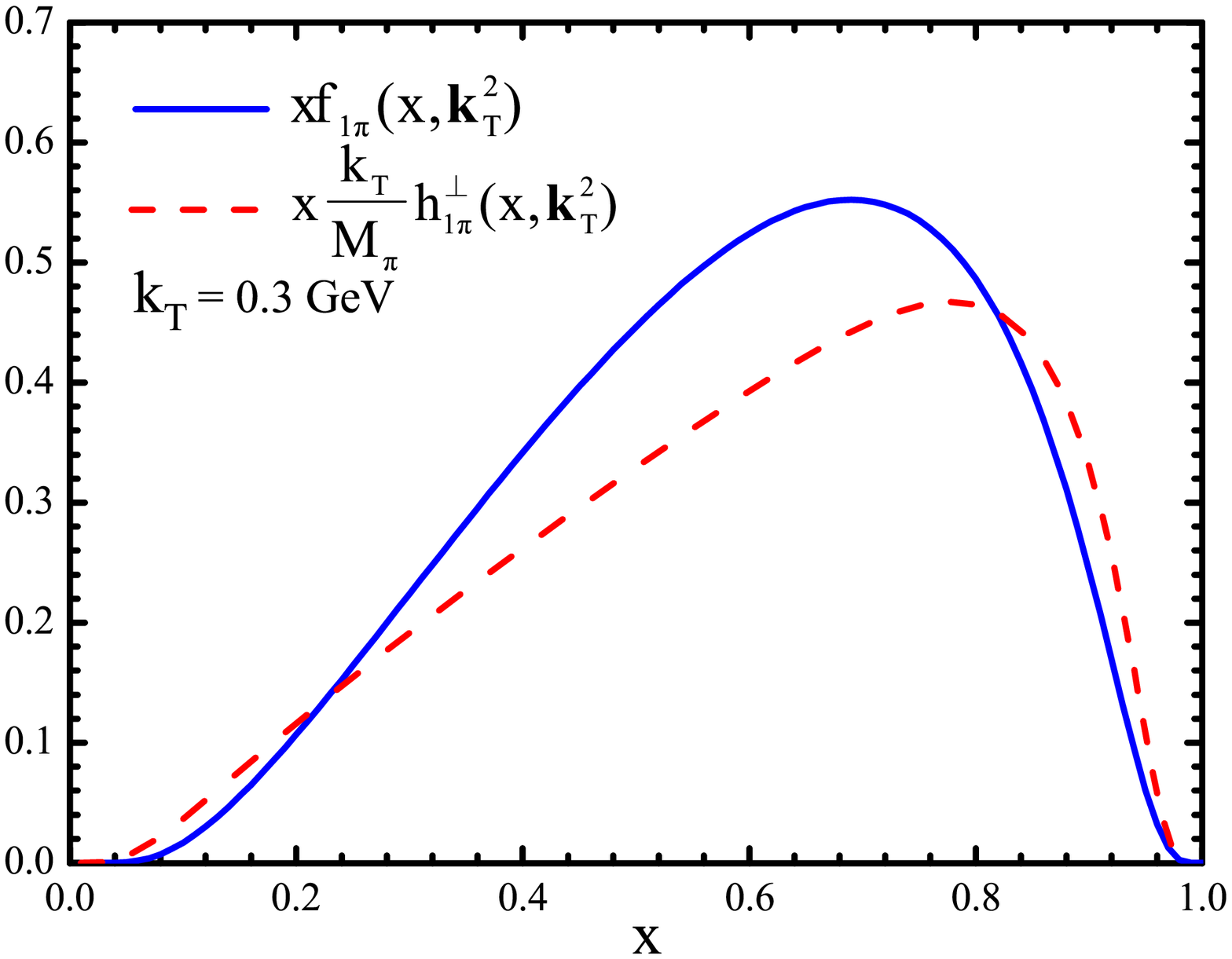}}~~~
\scalebox{0.41}{\includegraphics*[0pt,4pt][536pt,429pt]{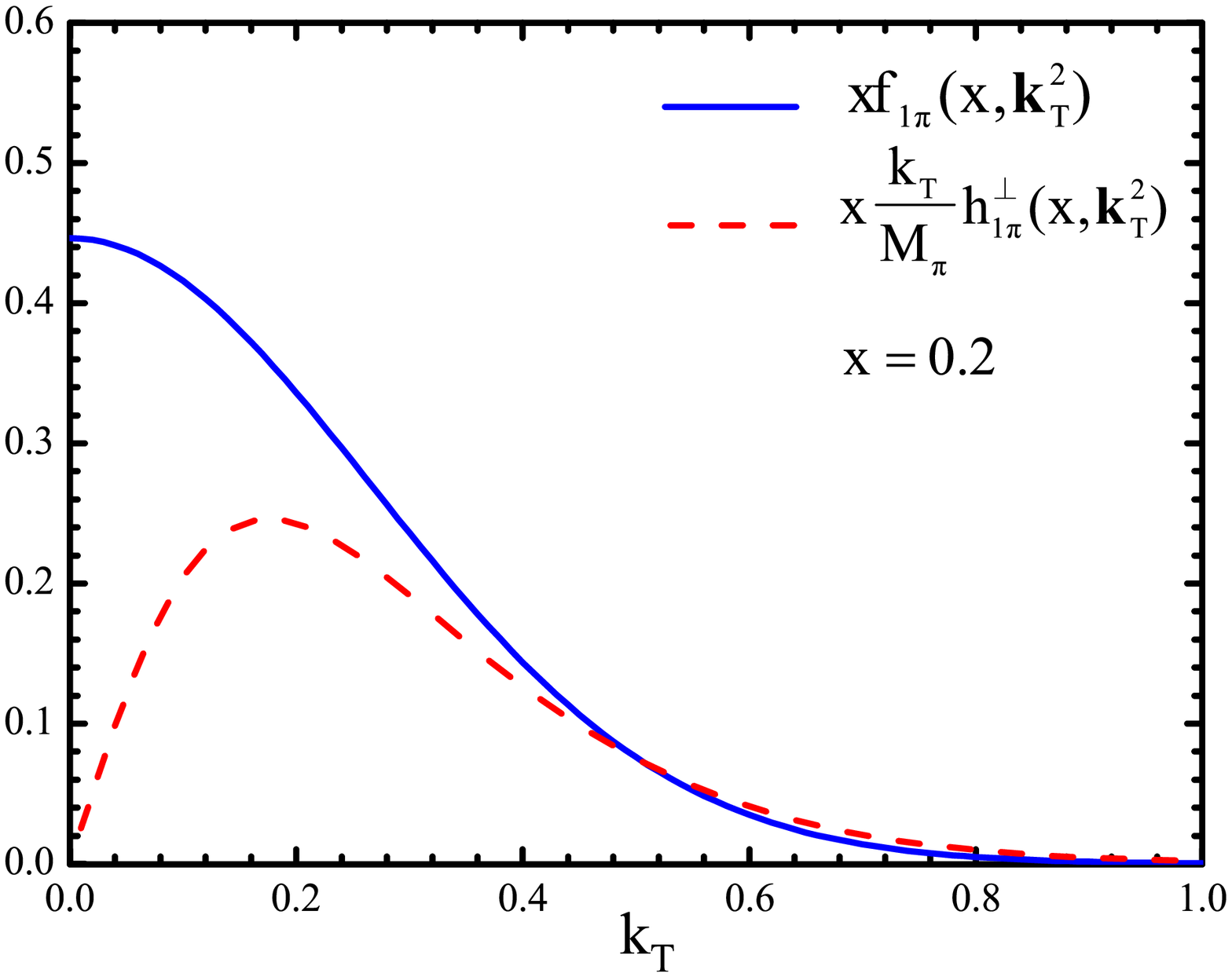}}
  \caption{Left panel: the $x$-dependence of  $\displaystyle{x{k_T \over M_\pi}h_{1\,\pi}^\perp(x,\bm k_T^2)}$ (dashed line) at $k_T=0.3\ \mathrm{GeV}$. Right panel: and $k_T$-dependence of $\displaystyle{x{k_T \over M_\pi}h_{1\,\pi}^\perp(x,\bm k_T^2)}$ (dashed line) at $x=0.2$. The solid lines in the two panels show the positivity bound $xf_1^q(x,\bm k_T^2)$.  }
  \label{fig:bmfunction}
\end{figure*}

In this section, we present our calculation on the Boer-Mulders function of the pion meson $h_{1\pi}^{\perp}(x,\bm k_T^2)$, using the overlap representation of the light-cone formalism.
Originally the overlap representation is applied to calculate various form factors of the nucleon~\cite{Brodsky:2000ii}, as well as the nucleon anomalous magnetic moment.
Recently it has also been adopted to calculate the quark Sivers function~\cite{Lu:2006kt,Bacchetta:2008af} and the quark Boer-Mulders function~\cite{Bacchetta:2008af}.
In the overlap representation, $h_{1\pi}^{\perp}$ may be calculated from the light-cone wave functions of the pion~\cite{Lu:2006kt,Bacchetta:2008af}:
\begin{align}
\label{eq:bmeq}
&\frac{(\hat{\bm{S}}_{qT}\times \bm{k}_T)\cdot\hat{\bm{P}}}{M_\pi}h^\perp_1(x,\bm{k}^2_T)=\int\frac{d^2\bm{k}^\prime_T}{16\pi^3}G(x,\bm{k}_T,\bm{k}^\prime_T)\nonumber\\
&\times \sum_{\bar{q}}[\psi^{\ast}_{\uparrow\lambda_{\bar{q}}}(x,\bm{k}_T)
\psi_{\uparrow\lambda_{\bar{q}}}(x,\bm{k}^\prime_T)-\psi^{\ast}_{\downarrow\lambda_{\bar{q}}}(x,\bm{k}_T)\nonumber\\
&\times \psi_{\downarrow\lambda_{\bar{q}}}(x,\bm{k}^\prime_T)]
+h.c.\,.
\end{align}
Here, $\uparrow$ and $\downarrow$ denotes the transverse polarization states of the struck quark,
The initial-state interactions (ISI) operator $G(x,\bm{k}_T,\bm{k}_T^{\prime})$ has the following form (in Drell-Yan process)~\cite{Lu:2006kt,Bacchetta:2008af}
\begin{align}
\mathrm{Im}\ G(x,\bm{k}_T,\bm{k}^\prime_T)=\frac{C_F\alpha_s}{2\pi}\frac{1}{(\bm{k}_T-\bm{k}^\prime_T)^2}.
\end{align}
It simulates a rescattering effect between the struck quark and the remnant of the pion via one gluon-exchange.

We insert the light-cone wave functions of the pion in Eq.~(\ref{eq:LCWFs}) into Eq.~(\ref{eq:bmeq}) to arrive at the expression
\begin{align}
\label{eq:bmf}
&\frac{\bm{k}^{2}_T}{M_\pi}{h}^{\perp}_{1\pi}(x,\bm{k}^{2}_T)={A}^{2}\int \frac{d^2\bm{k}^{\prime}_T}{16{\pi }^{3}}\frac{{C}_{F}{\alpha }_{s}}{2\pi }\frac{m}{(\bm{k}_T-\bm{k}^\prime_T)^2}\nonumber\\
&\times\frac{\bm{k}_T\cdot(\bm{k}_T-\bm{k}^\prime_T)}{\sqrt{({m}^{2}+\bm{k}^2_T)({m}^{2}+\bm{k}^{\prime 2}_T)}}\mathrm{exp}\left[-\frac{1}{8{\beta }^{2}}\frac{\bm{k}^{2}_T+\bm{k}^{\prime 2}_T+2{m}^{2}}{x(1-x)}\right].
\end{align}
Using the following integration formula:
\begin{align}
&\quad\int d^2\bm{k}^\prime_T \mathrm{exp}
(-a\bm{k}^{\prime2}_T)\frac{\bm{k}_T^2-\bm{k}_T\cdot \bm{k}_T^\prime}{(\bm{k}_T^\prime-\bm{k}_T)^2(\bm{k}_T^{\prime2}+b)^m}\\\nonumber
&=\pi\mathrm{exp}(ab)\left(\Gamma(1-m,ab)-\Gamma(1-m,a(\bm{k}^2_T+b))\right),
\end{align}
where $\Gamma(n,x)$ is the incomplete $\Gamma$-function
\begin{equation}
\Gamma(n,x)=\int^\infty_xdt\frac{e^{-t}}{t^{1-n}},
\end{equation}
we obtain the expression of the Boer-Mulders function $h^\perp_1(x,\bm{k}^2_T)$:
\begin{align}
\label{eq:BMFs}
&{h}^{\perp }_{1,\pi }(x,\bm{k}^{2}_T)=\frac{{C}_{F}{\alpha }_{s}}{16{\pi }^{3}}\frac{mM_\pi}{\sqrt{{m}^{2}+\bm{k}^{2}_T}}\frac{{A}^{2}}{\bm{k}^{2}_T}\mathrm{exp}[-\frac{1}{8{\beta }^{2}}\frac{\bm{k}^{2}_T+{m}^{2}}{x(1-x)}]\nonumber\\
&\times\left[\Gamma (\frac{1}{2},\frac{{m}^{2}}{8\beta^2 x(1-x)})-\Gamma (\frac{1}{2},\frac{\bm{k}^{2}_T+{m}^{2}}{8\beta^2 x(1-x)})\right].
\end{align}

Using the values of the parameters of $A$, $m$ and $\beta$ given in the previous section, we calculate the numerical result of $h^\perp_{1,\pi}(x,\bm{k}^2_T)$ at the model scale $\mu_0$.
For consistency, we adopt the value of the strong coupling at the model scale as $\alpha_s(\mu_0) = 0.911$.
In the left panel of Fig.~\ref{fig:bmfunction}, we plot the $x$-dependence of $\displaystyle{x{k_T \over M_\pi}h_{1\,\pi}^\perp(x,\bm k_T^2)}$~(dashed line) at a fixed transverse momentum $k_T=0.3\ \mathrm{GeV}$, while in the right panel we depict the $k_T$-dependence of $\displaystyle{x{k_T \over M_\pi}h_{1\,\pi}^\perp(x,\bm k_T^2)}$~(dashed line) at $x=0.2$.
The solid lines in the two panels show the positivity bound $xf_1^q(x,\bm k_T^2)$.

A theoretical constraint on the Boer-Mulders function is the positivity bound~\cite{Bacchetta:1999kz}
\begin{align}
{k_T \over M_\pi}\,h_{1\,\pi}^{\perp}(x, \bm k_T^2) \le f_{1\,\pi}(x,\bm k_T^2). \
\end{align}
From Fig.~\ref{fig:bmfunction} one can see that at the valence region $0.2<x<0.8$ and at $k_T<0.4$ GeV, the positivity bound for $h_{1\,\pi}^{\perp}(x, k_T^2)$ is satisfied.
Similar violation of the positivity bound was also observed in Ref.~\cite{Pasquini:2014ppa}.
An explanation was given in Ref.~\cite{Pasquini:2011tk}, stating that the violation may be due to the fact that T-odd TMD distributions (such as $h_1^\perp$) is evaluated to $\mathcal{O}(\alpha_s)$, while T-even TMD distributions (such as $f_1$) are truncated at $\mathcal{O}(\alpha_s^0)$ in model calculations.

\begin{figure}
  \includegraphics[width=1\columnwidth]{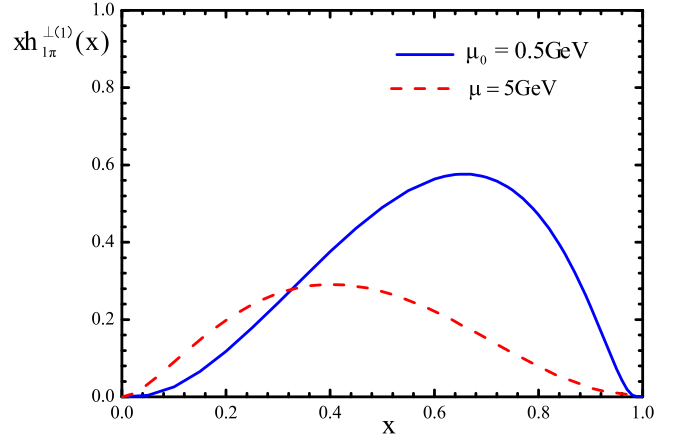}
  \caption{The $x$-dependence of $x{h}^{\perp(1)}_{1\,\pi}(x)$ at the model scale $\mu_0=0.5\ \mathrm{GeV}$~(solid line) compared with that at the scale $\mu=5\ \mathrm{GeV}$~(dashed line) after QCD evolution.}
  \label{fig:ktmoment}
\end{figure}

An important issue in phenomenological analysis is the evolution of TMDs.
Although much progress has been achieved, the exact evolution of the Boer-Mulders function is still unknown.
In order to gain the scale dependence of $h_{1\,\pi}^\perp$, we resort to the evolution of its first $\bm{k}_T$-moment, denoted by ${h}^{\perp (1)q}_{1\,\pi}(x)$, since ${h}^{\perp (1)q}_{1\,\pi}(x)$ appears in the transverse momentum weighted asymmetry naturally.
The first $\bm{k}_T$-moment of Boer-Mulders function in pion is defined as
\begin{align}
\label{eq:ktmoment}
{h}^{\perp (1)}_{1\,\pi}(x)=\int {d}^{2}\bm{k}_T\frac{\bm{k}_T^2}{2M_\pi^2}{h}^{\perp}_{1\,\pi}(x,\bm{k}_T^2).
\end{align}
At the tree level, ${h}^{\perp (1)q}_{1}(x)$ can be related to the twist-3 quark-gluon correlation function $T^{(\sigma)}_{q,F}(x,x)$
\begin{align}
{h}^{\perp (1)q}_{1}(x)\propto T^{(\sigma)}_{q,F}(x,x),
\end{align}
whereas the QCD evolution for $T^{(\sigma)}_{q,F}(x,x)$ is given in Ref.~\cite{Kang:2012em}.
For our purpose, we only consider the homogenous term in the evolution:
\begin{align}
P^{h_1^{\perp(1)}}_{qq}(x)\approx\Delta_TP_{qq}(x)-N_C\delta(1-x),\label{eq:evobm}
\end{align}
with $\Delta_TP_{qq}(x)=C_F\left[\frac{2z}{(1-z)_+}+\frac{3}{2}\delta(1-x)\right]$ being the evolution kernel for the transversity distribution function $h_1(x)$.
We note that the evolution kernel adopted in our calculation is different from the one in Ref.~\cite{Pasquini:2014ppa}, where $\Delta_TP_{qq}(x)$ is employed.

To perform the evolution of ${h}^{\perp (1)}_{1\pi}(x)$ numerically, we apply the {\sc{QCDNUM}} program~\cite{Botje:2010ay} and include the evolution kernel (\ref{eq:evobm}) in the code.
In Fig.~\ref{fig:ktmoment}, we plot the ${h}^{\perp (1)}_{1\,\pi}(x)$ vs $x$ at the model energy scale $\mu_0=0.5\mathrm{\ GeV}$ (solid line) and the evolved result at the energy scale $\mu=5\ \mathrm{GeV}$ (dashed line), respectively.
From the curves one can see that, with increasing energy scale, the evolution effect increases the size of ${h}^{\perp (1)}_{1\,\pi}(x)$ at smaller $x$ region and reduces the size at larger $x$ region, and it shifts the peak of the curve from higher $x$ region to lower $x$ region.

\section{$q_T$-Weighted $\cos 2 \phi$ asymmetry in the unpolarized $\pi^- p$ Drell-Yan at COMPASS}

\label{Sec:weighted}
In this section, we estimate the $q_T$-weighted $\cos 2\phi$ azimuthal asymmetry in the unpolarized $\pi^- p$ Drell-Yan process
\begin{equation}
\pi^-(P_\pi)+p(P) \longrightarrow\ell^+(l)+\ell^-(l')+X(P_X),
\end{equation}
using the quark distributions calculated in previous sections.

Usually the following invariants are defined to express the cross section of the Drell-Yan process:
\begin{align}
&x_1=\frac{Q^2}{2P_\pi\cdot q},~x_2=\frac{Q^2}{2P\cdot q},~ \tau=\frac{M^2}{s}=x_1x_2,\nonumber\\
&x_F=x_1-x_2,~y=\frac{1}{2}\ln \frac{q^+}{q^-}\overset{\mathrm{c.m.}}=\frac{1}{2}(1+\cos\theta),\nonumber
\end{align}
with $s=(P_\pi+P)^2$, and $Q^2=q^2=(l+l')^2$ the invariant mass squared of the lepton pair.

The angular differential cross section for unpolarized Drell-Yan process has the following general form
\begin{align}
\frac{1}{\sigma }\frac{d\sigma }{d\Omega }&=\frac{3}{4\pi }\frac{1}{\lambda +3}(1+\lambda\cos^2\theta +\mu \sin2\theta \cos\phi\nonumber\\
&+\frac{\nu }{2} \sin^2\theta \cos2\phi),
\label{eq:cross section}
\end{align}
where $\theta$ is the polar angle, and $\phi$ is the azimuthal angle of the hadron plane with respect to the dilepton plane.
The coefficients $\lambda$, $\mu$, $\nu$ in Eq.~(\ref{eq:cross section}) describe the sizes of different angular dependence.
In addition, $\nu$ stands for the asymmetry of the $\cos 2\phi$ azimuthal angular distribution of the dilepton.

According to the TMD factorization,
in the Collins-Soper frame~\cite{Collins:1977iv} the unpolarized Drell-Yan cross section at leading twist can be written as~\cite{Boer:1999mm}
\begin{align}
\label{eq:cs}
&\frac{d\sigma ({h}_{1}{h}_{2}\rightarrow l\bar{l}X)}{d\Omega d{x}_{1}d{x}_{2}{d}^{2}{\bm{q}}_{T}}=\frac{{\alpha }^{2}}{3{Q}^{2}}\sum _{q}\bigg{\{ }A(y)\mathcal{F}[{f}^{q}_{1}{f}^{\bar{q}}_{1}] +B(y)\cos2\phi\nonumber\\
& \times\mathcal{F}[(2\bm{\hat{h}}\cdot \bm{k}_{1T} \bm{\hat{h}}\cdot \bm{k}_{2T})-(\bm{k}_{1T}\cdot \bm{k}_{2T})]\frac{{h}^{\perp q}_1 {h}^{\perp \bar{q}}_1}{{M}_{1}{M}_{2}}\bigg{\}}
\end{align}
where we adopt the notation
\begin{align}
\mathcal{F}[\omega f\bar{f}]=&\int {d}^{2}\bm{{k}}_{1T}{d}^{2}\bm{{k}}_{2T}{\delta
}^{2}(\bm{k}_{1T}+\bm{{k}}_{2T}-\bm{{q}}_{T})\nonumber\\
&\times \omega f({x}_{1},\bm{{k}}_{1T}^2)\bar{f}({x}_{2},\bm{{k}}_{2T}^2)
\end{align}
to express the convolution of transverse momenta.
Here $\bm{q}_T$ is the transverse momentum of the lepton pair, and the unit vector $\bm{\hat{h}}$ is defined as $\bm{\hat{h}}=\frac{\bm{q}_{T}}{|\bm{q}_{T}|} = \frac{\bm{q}_{T}}{q_T}$.

The coefficients $A(y)$ and $B(y)$ in the c.m. frame of the lepton pair are as follows
\begin{align}
&A(y)=(\frac{1}{2}-y+y^2)=\frac{1}{4}(1+\cos^2 \theta),\nonumber\\
&B(y)=y(1-y)=\frac{1}{4} \sin^2 \theta. \nonumber
\end{align}

Comparing Eqs.~(\ref{eq:cross section}),~(\ref{eq:cs}), we can write the $\cos 2\phi$ asymmetry coefficient $\nu$ as ($\lambda=1,\ \mu=0$)
\begin{align}
\nu&=\frac{2\sum\limits_{q}\mathcal{F}\left[\left(2\bm{\hat{h}}\cdot \bm{k}_{1T} \bm{\hat{h}}\cdot \bm{k}_{2T}-\bm{k}_{1T}\cdot \bm{k}_{2T}\right){h}^{\perp \,q}_{1\,\pi}{h}^{\perp \,\bar{q} }_{1}\right]}{\sum\limits_{q}\mathcal{F}\left[{f}_{1}^qf_{1}^{\bar{q}}\right]
} ,\label{eq:nu}
\end{align}
with the summation running over all quark flavors.
It is understood that in the numerator and the denominator of Eq.~(\ref{eq:nu}), there are additional terms coming from the interchange $q\rightarrow \bar{q}$.

Another way to define azimuthal asymmetries is to use a weighting procedure~\cite{Bacchetta:2010si}£º
\begin{align}
A_{XY}^W\propto\frac{\langle W \rangle_{XY}}{\langle 1\rangle_{UU}}\equiv\frac{\int d\theta d\phi d\bm{q}_T^2 W d\sigma_{XY}}{\int d\theta d\phi d\bm{q}_T^2 d\sigma_{UU}},
\end{align}
where $d\sigma_{XY}$ denotes the cross section for hadrons with polarization states $X$ and $Y$, $W$ represents the weighting function which depends on the azimuthal angles and the transverse momentum of the dilepton $\bm{q}_T$.
To extract the $\cos 2\phi$ asymmetry, one can apply the weighting function as $W = \frac{\bm{q}_T^2}{4M_\pi M_P}\cos2\phi$:
\begin{align}
A^{q_T^2\cos 2\phi}_{UU}&=2\frac{{\langle\frac{\bm{q}_T^2}{4M_\pi M_P}\cos2\phi\rangle}_{UU}}{{\langle1\rangle}_{UU}}\nonumber \\
&\overset{\mathrm{TMD}}=2\frac{\sum\limits_q e_q^2 h_1^{\perp(1) q}(x_1)h_1^{\perp(1) \bar{q}}(x_2)}{\sum\limits_q e_q^2f_{1}^{q}(x_1)f_1^{\bar{q}}(x_2)},
\end{align}
where ${h}^{\perp(1)}_{1}(x)$ is the first $\bm{k}_T$-moment of Boer-Mulders function defined in Eq.~(\ref{eq:ktmoment}).

\begin{figure*}
  \centering
\scalebox{0.36}{\includegraphics*[45pt,45pt][710pt,520pt]{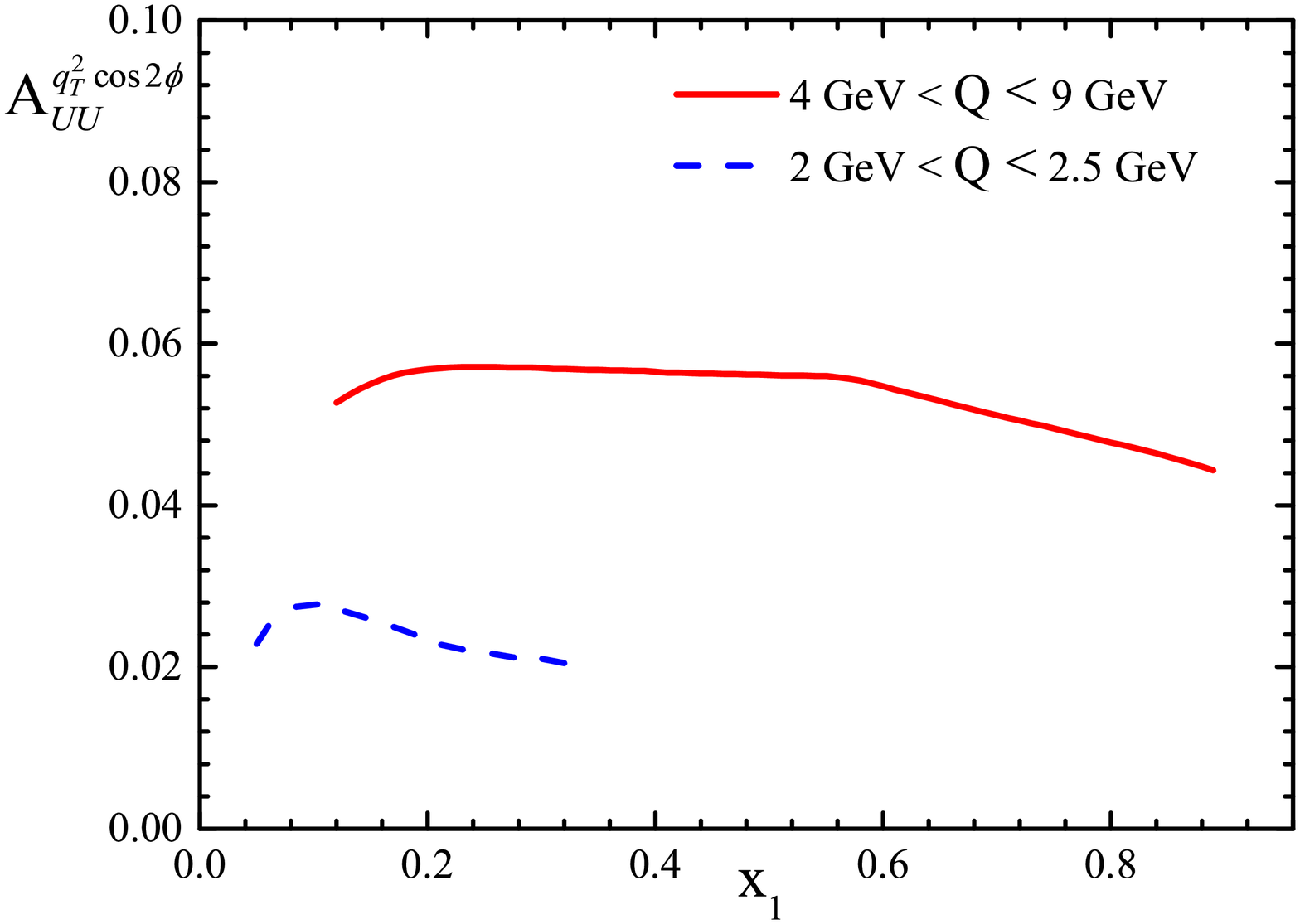}}~~~
\scalebox{0.36}{\includegraphics*[45pt,45pt][710pt,520pt]{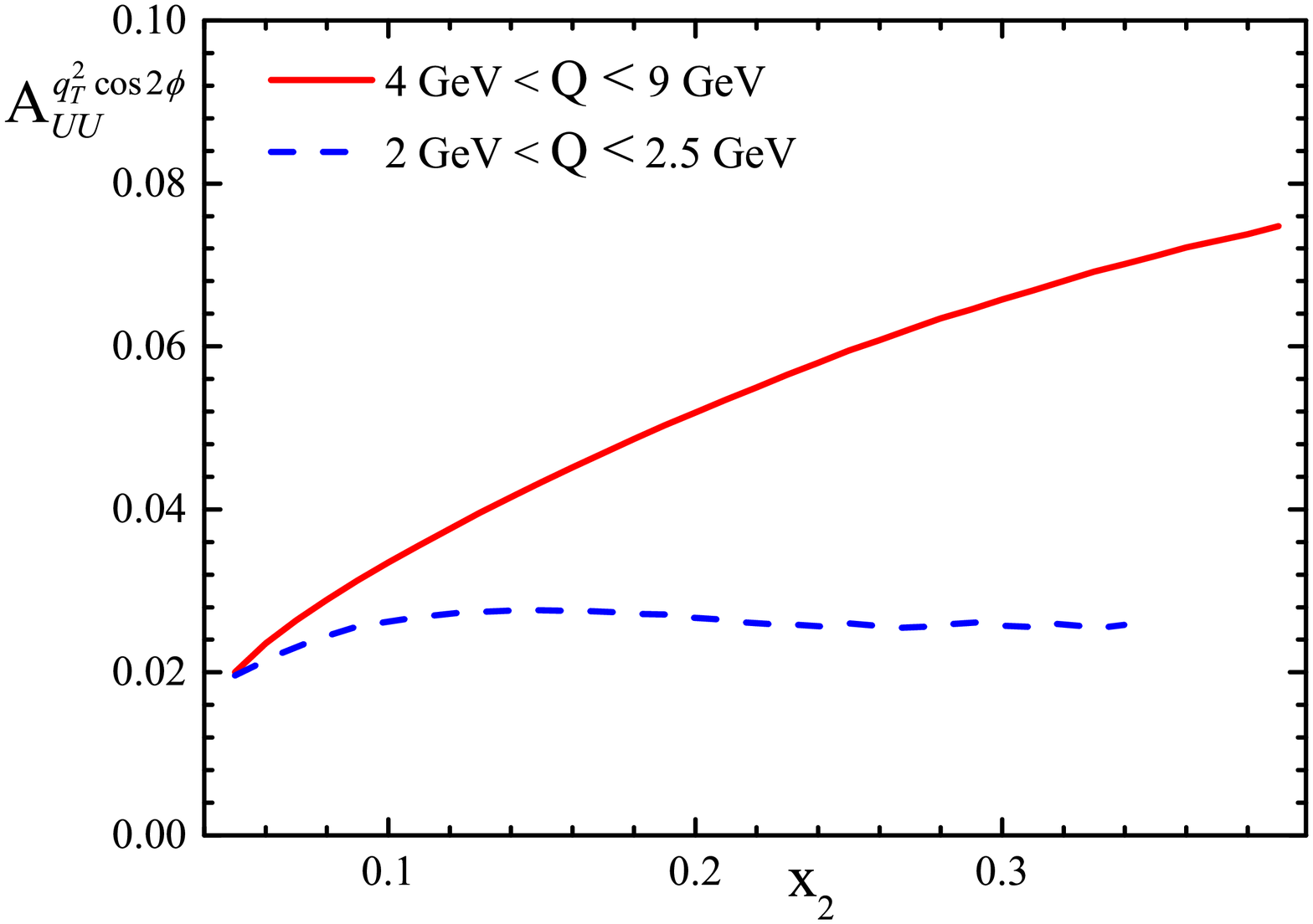}}
  \caption{The $x_1$-dependent (left panel) and $x_2$-dependent (right panel) $\frac{\bm{q}_T^2}{4M_\pi M_P}\cos2\phi$-weighted asymmetries in the unpolarized $\pi ^- p$ Drell-Yan process at the kinematics of COMPASS.}
  \label{fig:v_x12}
\end{figure*}

With all the ingredients mentioned above, we can express the $x_1$-dependent and $x_2$-dependent $q_T$-weighted $\cos 2\phi$ asymmetry as
 \begin{align}
 & A_{UU}^{q_T^2\cos 2\phi}(x_1) = \nonumber\\
 &\frac{2\displaystyle\int dx_2 \,\left[\sum\limits_q e^2_qh^{\perp(1)q}_{1\pi}(x_1)h^{\perp(1)\bar{q}}_{1}(x_2)
    \right]}
     {\displaystyle\int dx_1 \,\left[\sum\limits_q e^2_q f_{1\pi}^{q}(x_1)f_1^{\bar{q}}(x_2)\right]},\label{eq:x1}\\
 &A_{UU}^{q_T^2\cos 2\phi}(x_2) =\nonumber \\
 &\frac{2\displaystyle\int dx_1 \,\left[\sum\limits_q e^2_qh^{\perp(1)q}_{1\pi}(x_1)h^{\perp(1)\bar{q}}_{1}(x_2)
    \right]}
     {\displaystyle\int dx_1 \,\left[\sum\limits_q e^2_q f_{1\pi}^{q}(x_1)f_1^{\bar{q}}(x_2)\right]}
 . \label{eq:x2}
 \end{align}
Here, we use $f_1^q(x)$ to denote the unpolarized distribution of the proton, and we use $h_{1}^{\perp (1)}(x)$ to denote the first $\bm k_T$-moment of the Boer-Mulders function of the proton.
For $f_1^q(x)$, we adopt the leading-order set of the MSTW2008 parametrization~\cite{Martin:2009iq}; as for
Boer-Mulders function of the proton needed in the calculation, we adopt the parametrization in Ref.~\cite{Lu:2009ip}:
\begin{align}
&{h}^{\perp q}_1(x,\bm{k}^2_T)={h}^{\perp q}_{1}(x)\frac{\exp (-\bm{k}_T^2/k_{bm}^2)}{\pi k_{bm}^2},\\
&{h}^{\perp q}_{1}(x)=H_{q}{x}^{c^{q}}(1-x)^{b}{f}^{q}_{1}(x).
\label{eq:bmproton}
\end{align}

Using Eqs.~(\ref{eq:x1}) and (\ref{eq:x2}) and including the evolution effects of the distributions, we calculate the weighted asymmetry $A_{UU}^{q_T^2\cos2\phi}$ at the kinematics of COMPASS, at which the unpolarized Drell-Yan data may be taken with a $\pi^-$ beam of 190 GeV colliding on a $\mathrm{NH}_3$ proton target.
The available kinematical region is as~\cite{Adolph:2016dvl}
\begin{align}
 &0.05<x_1<0.9,\quad  0.05<x_2<0.4.\nonumber
\end{align}
Besides, there are two $Q$ ranges can be selected for the analysis of the DY data: the intermediate mass range $2\ \mathrm{GeV} < Q < 2.5\ \mathrm{GeV}$ and the high mass range $4\ \mathrm{GeV} < Q < 9\ \mathrm{GeV}$.

In the left and right panels of Fig.~\ref{fig:v_x12}, we plot the $x_1$-dependent and $x_2$-dependent weighted asymmetry $A_{UU}^{q_T^2\cos2\phi}$.
The solid lines and dashed lines depict the asymmetries at the high mass region and the intermediate mass region, respectively.
We find that the weighted $\cos 2\phi$ asymmetries contributed by the Boer-Mulders function are sizable, about several percent.
Particularly, the asymmetry at the high mass region is larger than that at the intermediate mass region.
This is because the asymmetry is dominated by the contribution of the Boer-Mulders function in the valence region.
Therefore, it is promising to measure the weighted $\cos2\phi$ asymmetry at COMPASS to shed light on the pion Boer-Mulders function in the valence region.
It is also found that at the high mass region the asymmetry increase with increasing $x_2$ in the region $x_2 <0.4$.
Thus, the $x_2$-dependence of the asymmetry may be tested at the kinematics of COMPASS.

\begin{figure*}
  \centering
  \scalebox{0.36}{\includegraphics*[50pt,50pt][720pt,520pt]{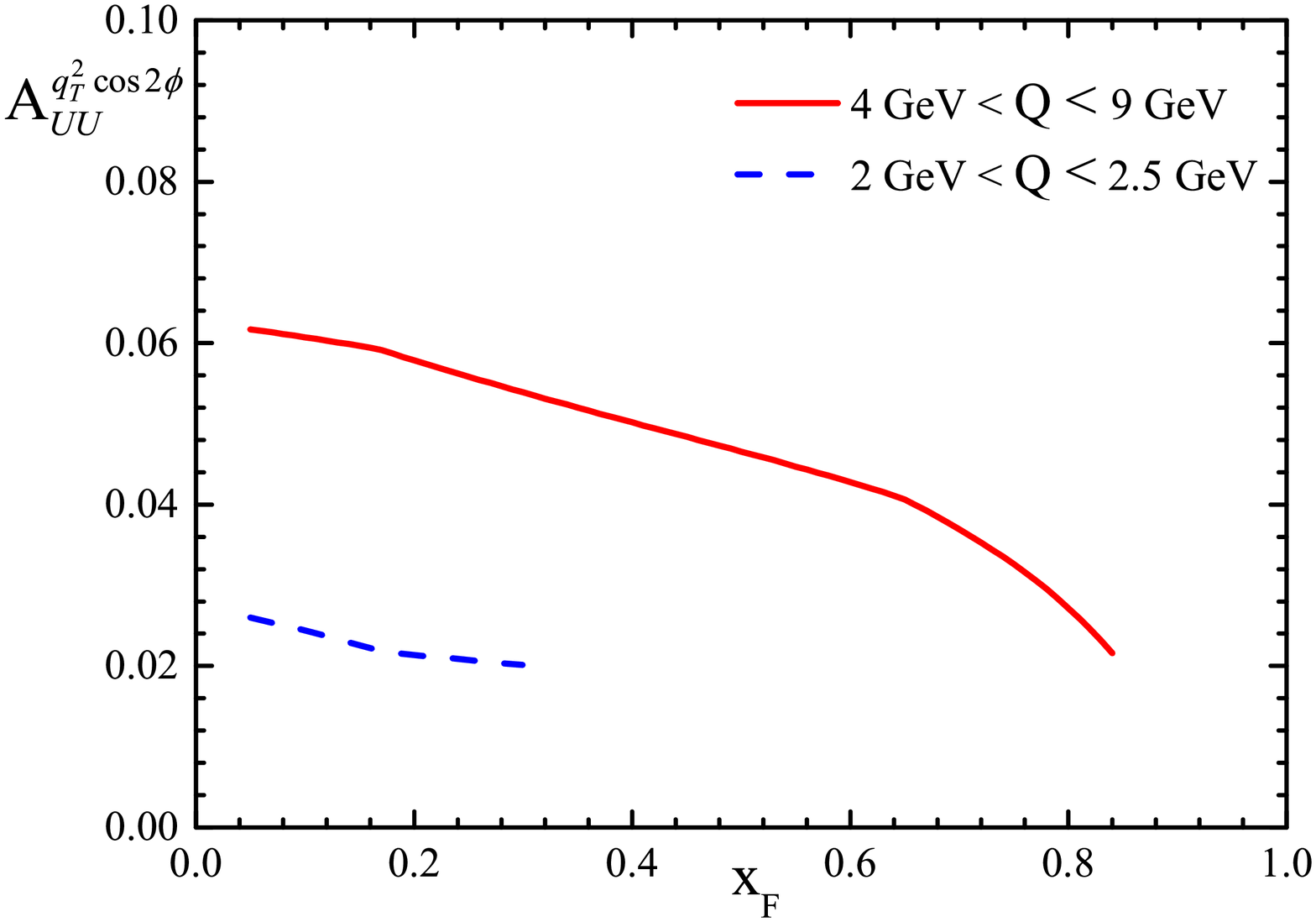}}~~~
\scalebox{0.36}{\includegraphics*[50pt,50pt][720pt,520pt]{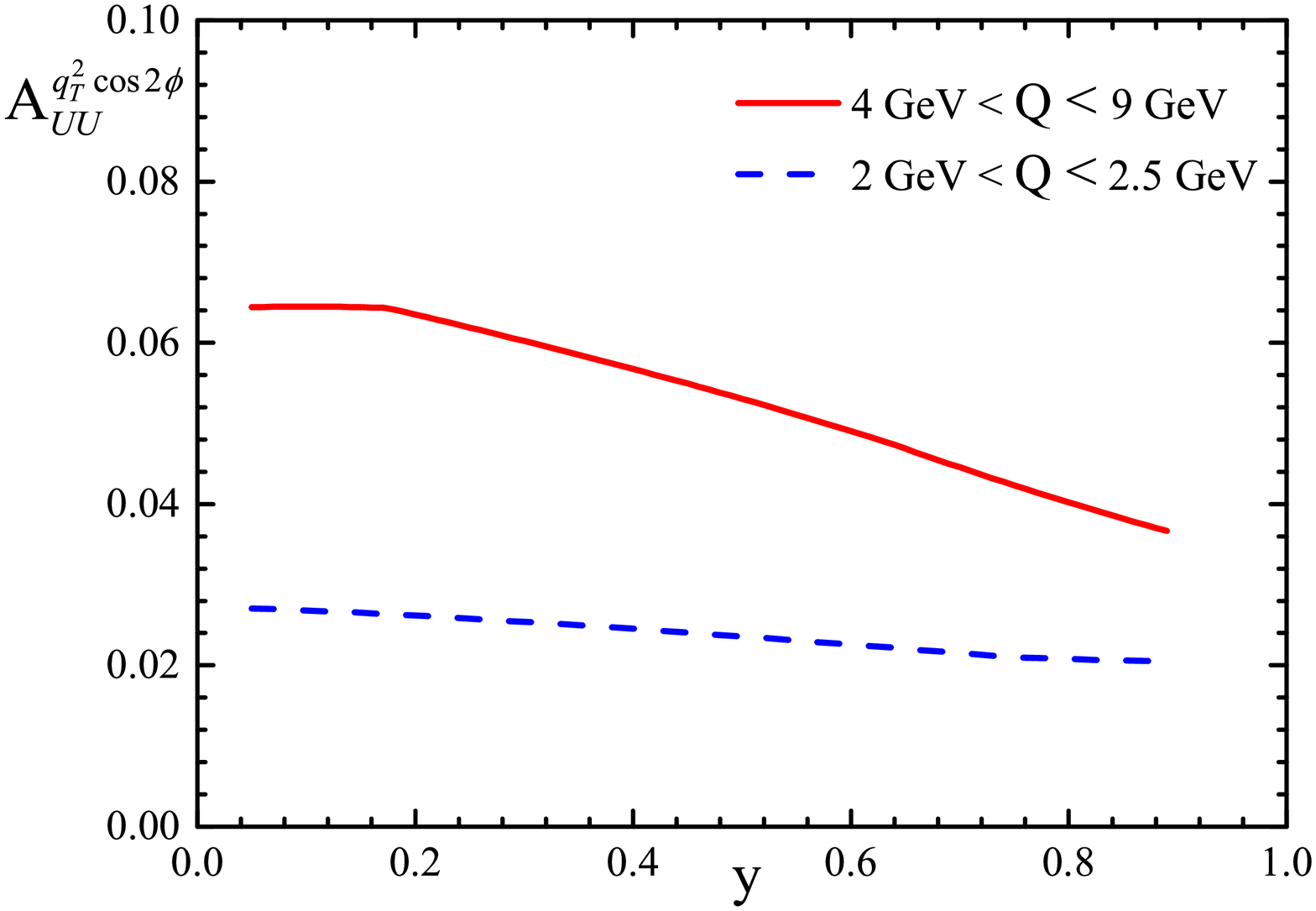}}
  \caption{Similar to Fig.~\ref{fig:v_x12},but for the $x_F$-dependent (left panel) and the $y$-dependent (right panel) asymmetries.}
  \label{fig:v_xfy}
\end{figure*}

Finally, we make further prediction on the $q_T$-weighted $\cos2\phi$ asymmetry vs $x_F$ and $y$ at COMPASS.
To do this we apply the relation among $x_1$, $x_2$ and $x_F$:
 \begin{align}
 &x_1=\frac{x_{F}+\sqrt{{x}^{2}_{F}+4\tau}}{2},\nonumber
 &x_2=\frac{-x_{F}+\sqrt{{x}^{2}_{F}+4\tau}}{2},\nonumber
 \end{align}
 with $\tau = Q^2/s$.
Therefore, the $x_F$-dependent $\cos 2\phi$ weighted asymmetry can be expressed as
\begin{align}
 &A_{UU}^{q_T^2\cos 2\phi}(y) =\nonumber\\
 &\frac{2\displaystyle\int dQ^2 \,J_1\,\left[\sum\limits_q e^2_qh^{\perp(1)q}_{1\pi}(x_1)h^{\perp(1)\bar{q}}_{1}(x_2)\right]}
     {\displaystyle\int dQ^2 \,J_1\,\left[\sum\limits_q e^2_q f_{1\pi}^{q}(x_1)f_1^{\bar{q}}(x_2)\right]},
 \end{align}
where $\displaystyle J_1=\frac{1}{s\sqrt{x_F^2+4\tau}}$.

Similarly, we can express $x_1,x_2$ in terms of $y,Q^2$ as
\begin{align}
\label{eq:x12y}
&x_1=\frac{Q}{\sqrt{s}}e^{y}, ~~~
x_2=\frac{Q}{\sqrt{s}}e^{-y} \nonumber
\end{align}
to define the $y-$dependent $\cos2\phi$ weighted asymmetry:
 \begin{align}
 &A_{UU}^{q_T^2\cos 2\phi}(y) =\nonumber\\
 &\frac{2\displaystyle\int dQ^2 \,{1\over s}\,\left[\sum\limits_q e^2_qh^{\perp(1)q}_{1\pi}(x_1)h^{\perp(1)\bar{q}}_{1}(x_2)
    \right]}
     {\displaystyle\int dQ^2 \,{1\over s}\,\left[\sum\limits_q e^2_q  f_{1\pi}^{q}(x_1)f_1^{\bar{q}}(x_2)\right]}.
 \end{align}

In the left and right panels of Fig.~\ref{fig:v_xfy}, we plot the $x_F$ and $y$ dependent asymmetries at the kinematical region of COMPASS, respectively.
We find that in this case the asymmetries decrease with increasing $x_F$ and $y$.

Finally, to study the dependence of the asymmetry on the proton
Boer-Mulders function, we also implement two different choice of the proton Boer-Mulders function to estimate the weighted asymmetry in unpolarized $\pi^- p$ Drell-Yan at the kinematics of COMPASS.
The first one is the proton Boer-Mulders function extracted in
Ref.~\cite{Barone:2009hw}, the second one is the model result using light-cone wave functions of the proton from Ref.~\cite{Bacchetta:2008af}.
We find that the asymmetry calculated from these two sets of the proton Boer-Mulders function is one or two times larger than the asymmetry shown in Figs.~\ref{fig:v_x12} and \ref{fig:v_xfy}, although the sign of the asymmetry is consistent with our result.
This is because that the proton Boer-Mulders function either in Ref.~\cite{Barone:2009hw} or in Ref.~\cite{Bacchetta:2008af} is larger than the proton Boer-Mulders function extracted in Ref.~\cite{Lu:2009ip}.
Our study shows that the asymmetry strongly depends on the choice of the proton Boer-Mulders function.
Therefore, the measurement of the at COMPASS should also provide constraints on the the proton Boer-Mulders function.

\section{Conclusion}

\label{Sec:conclusion}

In this work, we studied the transverse momentum weighted $\cos2\phi$ asymmetry contributed by the Boer-Mulders function in the unpolarized $\pi^- p$ Drell-Yan process.
To do this we calculated the leading twist TMD distributions of the pion meson using the light-cone formalism.
In particular, we applied the light-cone wave functions derived from the minimal Fock-state of the pion to compute the unpolarized distribution $f_{1\pi}(x,\bm{k}_T^2)$ and the Boer-Mulders function $h_{1\pi}^\perp(x,\bm{k}^2_T)$ for valence quarks.
The Boer-Mulders function of the proton needed in the calculation was taken from the available parametrization.
We estimated the $q_T$-weighted $\cos2\phi$ asymmetry in the unpolarized $\pi^- p$ Drell-Yan process at the kinematics of COMPASS.
In the study we also included the QCD evolution effect of the unpolarized distribution function $f_{1\pi}(x)$ as well as that of the first $\bm{k}_T$-moment of the Boer-Mulders function $h_{1\pi}^{\perp(1)}(x)$.
The numerical calculation result shows that the asymmetries are all sizable, about several percent.
Therefore, there is a great opportunity to access the $q_T$-weighted $\cos 2\phi$ asymmetry in the unpolarized $\pi^- p$ Drell-Yan process at COMPASS.
We took into account two different $Q$ ranges, the intermediate mass and high mass region, to present the results for comparison.
We find that the asymmetry at the high mass region is ideal to obtain the information of the pion Boer-Mulders function, especially in the valence region.

\section*{Acknowledgements}
This work is partially supported by the National Natural Science
Foundation of China (Grants No.~11575043, and No.~11120101004), and by the Qing Lan Project. X.Wang is supported by the Scientific Research Foundation of Graduate School of Southeast University (Grants No.~YBJJ1667).

\end{document}